\documentstyle[12pt,epsfig]{article}
\begin{document}
\topskip 2cm
\begin{titlepage}
\rightline{ \large{ \bf December 1999} }
\rightline{ \large{ \bf hep-ph/9912501} }
\begin{center}
{\large\bf The Pseudothreshold Expansion of  the 2-loop Sunrise Selfmass
         Master Amplitudes.} \\
\vspace{2.5cm}
\begin{center}
{\large {\bf
M.~Caffo$^{ab}$,
H.~Czy{\.z}\ $^{c}$} and {\bf E.~Remiddi$^{ba}$ \\ } }
\end{center}

\begin{itemize}
\item[$^a$]
             {\sl INFN, Sezione di Bologna, I-40126 Bologna, Italy }
\item[$^b$]
             {\sl Dipartimento di Fisica, Universit\`a di Bologna,
             I-40126 Bologna, Italy }
\item[$^c$]
             {\sl Institute of Physics, University of Silesia,
             PL-40007 Katowice, Poland }

\end{itemize}
\end{center}

\noindent
e-mail: {\tt caffo@bo.infn.it \\
\hspace*{1.3cm} czyz@us.edu.pl \\
\hspace*{1.3cm} remiddi@bo.infn.it \\ }
\vspace{.5cm}
\begin{center}
\begin{abstract}
  The values at pseudothreshold of two loop sunrise master amplitudes
  with arbitrary masses
  are obtained by solving a system of differential equations.
  The expansion at pseudothreshold of the amplitudes is constructed
  and some lowest terms are explicitly presented.
\end{abstract}
\end{center}
\scriptsize{ \noindent ------------------------------- 
}
\vfill
\end{titlepage}
\pagestyle{plain} \pagenumbering{arabic}
\newcommand{\F}[1]{F_#1(n,m_1^2,m_2^2,m_3^2,p^2)}
\newcommand{\B}[2]{#1_#2(n,m_1^2,m_2^2,m_3^2,p^2)}
\newcommand{\D}{D(m_1^2,m_2^2,m_3^2,p^2)}
\newcommand{\A}[3]{#1_{#2,#3}(n,m_1^2,m_2^2,m_3^2,p^2)}
\newcommand{\HH}[1]{H^{(#1)}(n,m_1,m_2,m_3)}
\newcommand{\co}{\left(p^2+(m_1+m_2-m_3)^2\right)}
\newcommand{\GG}[1]{G_#1(n,m_1,m_2,m_3)}
\def\Li2{\hbox{Li}_2}
\def\LLL{L(m_1^2,m_2^2,m_3^2)}
\def\a{\alpha}
\def\app{{\left(\frac{\alpha}{\pi}\right)}}
\newcommand{\Eq}[1]{Eq.(\ref{#1})}
\newcommand{\labbel}[1]{\label{#1}}
\newcommand{\cita}[1]{\cite{#1}}
\newcommand{\dnk}[1]{ \frac{d^nk_{#1}}{(2\pi)^{n-2}} }
\newcommand{\e}{{\mathrm{e}}}
\newcommand{\verso}[1]{ {\; \buildrel {n \to #1} \over{\longrightarrow}}\; }
\section{Introduction.}
The sunrise graph (also known as sunset or London transport diagram)
appears naturally, as a consequence of tensorial reduction,
in several higher order calculations in gauge theories. Due
to the presence of heavy quarks, vector bosons and Higgs particles
all the internal lines may carry a different mass, so that
sunrise amplitudes depend in the general case on three internal masses
\( m_i,\ i=1,2,3, \) and the external scalar variable \( p^2 \), if
\( p_\mu \) is the external momentum (in \(n\)-dimensional Euclidean space).

For a proper understanding of their behaviour as well as for a check
of the numerical calculations, it is convenient to know the amplitudes
off-shell, and also around some particular values of \( p^2 \),
such as \( p^2 =0\), \( p^2 =\infty\), at the threshold and at the
pseudothresholds.

This paper is devoted to the analytic evaluation of the coefficients
of the expansion of the sunrise amplitudes in \( p^2 \), at the
pseudothreshold value \( \ \  p^2 = -(m_1+m_2-m_3)^2 \ \  \) \linebreak
( the other pseudothreshold values
\( -(m_1-m_2+m_3)^2, \) \( -(m_1-m_2-m_3)^2, \)
can be easily obtained by permutation of the masses );
the approach relies on the exploitation of the information contained
in the linear system of first order differential equations in \( p^2 \),
which is known to be satisfied by the sunrise amplitudes themselves
\cita{CCLR}. It is to be noted that all the above points
( \( p^2 =0,\ \infty\ ,\) threshold and pseudothresholds) correspond
to the Fuchsian points of the differential equations,
which therefore emerge as a natural tool for their discussion.

The analytic properties of Feynman diagrams at threshold and
pseudothresholds are well known, see for example \cita{IZ}.
The sunrise diagram, with different masses, has been
investigated in \cita{BBBS}, while in \cita{BDU} the values of the amplitudes
at threshold and pseudothreshold were obtained.
With the method established in \cita{S} and \cita{BS}, further, the
expansion around threshold was recently obtained in \cita{DS}.

The sunrise amplitudes are regular at any of the pseudothresholds,
say \( s_0 \), so that they can be expanded as a single power series in
\( (p^2-s_0) \) around that point \cita{I}.
When the expansions are inserted in the differential equations, the
equations become
a set of algebraic equations in the coefficients of the expansions;
the obtained algebraic equations can then be solved recursively,
for arbitrary value of the dimension \( n \),
once the initial conditions, {\it i.e.} the values of the sunrise
amplitudes at the considered pseudothreshold, are given.
As those initial values are in turn functions of the masses, we find
in our approach a system of linear differential equations in the masses
satisfied by them.
We expand the equations in the masses in the dimension \( n \)
around \( n=4 \) and solve them explicitly up to the finite part in
\( (n-4) \).
This result for \( n=4 \) and \( p^2=s_0 \) is in agreement with the
literature \cita{BDU}.

We then look at the recursive solution of the algebraic equations for
the coefficients of the expansion in \( (p^2-s_0) \), not yet discussed in
the literature for the pseudothreshold (our results are given in the
Sections 4 and 5). It turns out that the
formula, expressing the first order coefficients of the \( (p^2-s_0) \)
expansion in terms of the zeroth order values, involves the coefficient
\( 1/(n-4) \).  Therefore the finite
part at \( (n-4) \) of the first order terms in the \( (p^2-s_0) \) expansion
involve the first order terms in \( (n-4) \) of the zeroth order terms
in \( (p^2-s_0) \). Rather than evaluating those first order terms in
\( (n-4) \), we prefer to evaluate directly the required finite part in
\( n \) of the first order terms of the \( (p^2-s_0) \) expansion by means
of a suitably subtracted dispersion relation in \( (p^2-s_0) \) at \( n=4 \).
Fortunately, the formulae expressing the higher order coefficients of
the \( (p^2-s_0) \) expansion involve coefficients like \( 1/(n-5), 1/(n-6) \)
etc., which are finite at \( n=4 \), and can be used without further
problems for evaluating those higher order terms.

The plan of the paper is as follows. In the second section the system
of differential equations for the sunrise master amplitudes is recalled;
in the third section the resulting differential equations in the masses
for the values at the pseudothreshold are derived, expanded in \( (n-4) \)
and solved explicitly up to the finite part in \( (n-4) \); in the fourth
section the expansion in \( p^2 \) at the pseudothreshold is discussed;
in the fifth section the finite part in \( (n-4) \) of the first term in
the \( p^2 \) expansion is given. The sixth section is a short summary,
while the Appendix contains a list of various functions (mainly
polynomials) used in the text.
\section{The equations.}
It is known that the two-loop sunrise self-mass graph with arbitrary masses
\( m_1, m_2, m_3 \) has four independent master amplitudes \cita{Tarasov},
which are referred to, as in \cita{CCLR}, by
\begin{equation}
      \F{\alpha} , \hspace{1truecm} \alpha=0,1,2,3 \ ,
\labbel{Falpha}\end{equation}
where \( n \) is the continuous number of dimensions, \( m_i, i=1,2,3 \)
the three masses and \( p_\mu \) the external \(n\)-momentum.
\( \F{0} \) is the scalar amplitude
\begin{eqnarray}
  \F{0} &=& \nonumber \\
      && {\kern-200pt} \int \dnk{1} \int \dnk{2} \;
      \frac{ 1 }
           { (k_1^2+m_1^2) (k_2^2+m_2^2) ( (p-k_1-k_2)^2+m_3^2 ) } \ ,
\labbel{F0} \end{eqnarray}
 while for \( i=1,2,3 \)
\begin{equation}
      \F{i} = - \frac{\partial}{\partial m_i^2} \F{0} \ ,
\labbel{Fi}\end{equation}
but the 4 amplitudes are otherwise independent ({\it i.e.} none of them
can be expressed as a linear combinations of the others times rational
functions of the arguments). \par
In \cita{CCLR} it was shown how to write for them a system
of first order linear differential equations of the form
\begin{eqnarray}
 p^2\frac{\partial}{\partial p^2}\F{0} {\kern-10pt}&=& {\kern-10pt}(n-3)\F{0}
           +\sum_{i=1}^3 m_i^2 \F{i} \ , \nonumber\\
 p^2\frac{\partial}{\partial p^2}\F{i} &=& \frac{1}{\D} \nonumber\\
          && {\kern-130pt}\biggl[ \sum_{\alpha=0}^3 \A{A}{i}{\alpha}\F{\alpha}
                     + \B{B}{i} \biggr] \ ,
\labbel{Feqs}\end{eqnarray}
where the \( \A{A}{i}{\alpha} \) are known polynomials of the arguments,
and the functions
\( \B{B}{i} \), constituting the non homogeneous part of the equations, are
also known, being the combinations of other known polynomials times
the products of two one-denominator, one-loop vacuum amplitudes of two
different masses, \( T(n,m_i^2)T(n,m_j^2) \), where, as in \cita{CCLR},
\begin{equation}
  T(n,m^2) = C(n) \frac{m^{n-2}}{(n-2)(n-4)} \ ,
\labbel{Tn}\end{equation}
and \( C(n) \) is a coefficient depending on \( n \) which at \( n=4 \)
takes the value \( C(4) = 1 \) (its explicit form, which is essentially
irrelevant in practice, can be found in \cita{CCLR}).
                                                       \par
The explicit form of the \( \A{A}{i}{\alpha}, \B{B}{i} \), appearing
in \Eq{Feqs} is given in \cita{CCLR} and
is not repeated here for the sake of brevity; for the following, it
is sufficient to give the explicit expression of the polynomial \( \D \),
which reads
\begin{eqnarray}
   \D =&& \left[ p^2+(m_1+m_2+m_3)^2 \right]
          \left[ p^2+(m_1+m_2-m_3)^2 \right]   \nonumber\\
      &&  \left[ p^2+(m_1-m_2+m_3)^2 \right]
          \left[ p^2+(m_1-m_2-m_3)^2 \right] \ .
\labbel{D}\end{eqnarray}
The above polynomial vanish when \( p^2 \) takes one of the three
pseudothreshold values
\( -(m_1+m_2-m_3)^2, \) \( -(m_1-m_2+m_3)^2, \) \( -(m_1-m_2-m_3)^2, \)
and at the physical threshold \( p^2 = -(m_1+m_2+m_3)^2 \).
It is known that the \( \F{i} \) and their \( p^2 \)-derivatives are
regular at the pseudothresholds; therefore, the apparent pseudothreshold
pole in the {\it r.h.s.} of \Eq{Feqs} must be canceled by a corresponding
zero in the numerator. \par
The recurrence relations solving the integration by part identities of
\cita{ChetTka} for the sunrise amplitudes can be also written (see
\cita{Tarasov} or the Appendix of \cita{CCLR}) in a form similar to \Eq{Feqs},
such as
\begin{eqnarray}
 \frac{\partial}{\partial m_3^2}\F{i} &=& \frac{1}{\D} \nonumber\\
          && {\kern-150pt}\biggl[ \sum_{\alpha=0}^3 \A{R}{i}{\alpha}\F{\alpha}
                     + \B{S}{i} \biggr] \ ,
\labbel{Frec}\end{eqnarray}
where \( i=1,2,3 \), the \( \A{R}{i}{\alpha} \) and the \( \B{S}{i} \)
have the same structure as the \( \A{A}{i}{\alpha} \) and the
\( \B{B}{i} \) of
\Eq{Feqs}; as in \Eq{Feqs}, the apparent pseudothreshold singularity of
\Eq{Frec} must be canceled by a corresponding zero of the numerator.
\par
\section{The pseudothreshold values of the master amplitudes.}
As the master amplitudes are regular at the
pseudothresholds, considering for definiteness the pseudothreshold
at \( p^2 = -(m_1+m_2-m_3)^2 \), we can expand \( \F{0} \) in \( p^2 \)
around that point as
\begin{eqnarray}
 \F{\alpha} &=& \HH{\alpha,0} \nonumber\\
       &+& \HH{\alpha,1} \co \nonumber\\
       &+& \HH{\alpha,2} \co^2 + ... , \nonumber \\
 &&\alpha=0,1,2,3 \ ,
\labbel{expa}\end{eqnarray}
from which we can write at once, by using the definitions \Eq{Fi},
\begin{eqnarray}
 \GG{0} &\equiv& F_{0}\left(n,m_1^2,m_2^2,m_3^2,p^2=-(m_1+m_2-m_3)^2\right)
      \nonumber\\ &=& \HH{0,0} \ , \nonumber\\
 \GG{1} &\equiv& F_{1}\left(n,m_1^2,m_2^2,m_3^2,p^2=-(m_1+m_2-m_3)^2\right)
      \nonumber\\ &=& \HH{1,0}
      \nonumber\\ &&{\kern-50pt}= - \frac{\partial}{\partial m_1^2} \HH{0,0}
        - \frac{m_1+m_2-m_3}{m_1}\HH{0,1} \ , \nonumber\\
 \GG{2} &\equiv& F_{2}\left(n,m_1^2,m_2^2,m_3^2,p^2=-(m_1+m_2-m_3)^2\right)
      \nonumber\\ &=& \HH{2,0}
      \nonumber\\ &&{\kern-50pt}= - \frac{\partial}{\partial m_2^2} \HH{0,0}
        - \frac{m_1+m_2-m_3}{m_2}\HH{0,1} \ , \nonumber\\
 \GG{3} &\equiv& F_{3}\left(n,m_1^2,m_2^2,m_3^2,p^2=-(m_1+m_2-m_3)^2\right)
      \nonumber\\ &=& \HH{3,0}
      \nonumber\\ &&{\kern-50pt}= - \frac{\partial}{\partial m_3^2} \HH{0,0}
        + \frac{m_1+m_2-m_3}{m_3}\HH{0,1} \ . \nonumber\\
\labbel{thrval}\end{eqnarray}
By inserting the above expansions in the four equations \Eq{Feqs} and in the
three equations \Eq{Frec} for \( i=1,2,3 \), we find only five
algebraically independent equations for \( \GG{\alpha},\alpha=0,1,2,3 \)
and their derivatives in respect to \( m_3 \), which can be written as

\newcommand{\G}[1]{G_{#1}(n,m_1,m_2,m_3)}
\newcommand{\GP}[1]{ G^{#1}_0(m_1,m_2,m_3)}

\begin{eqnarray}
 &&{\kern-10pt}- \frac{(n-3)(3n-8)}{2}\G{0} = \nonumber \\
 &&{\kern-10pt}+ (n-3) m_1 (2m_1 -m_3+m_2) \G{1}
 + \frac{(n-2)^2}{4m_2m_3} T(n,m_2^2) T(n,m_3^2)\nonumber \\
 &&{\kern-10pt}+ (n-3) m_2 (2m_2 -m_3+m_1) \G{2}
 + \frac{(n-2)^2}{4m_1m_3} T(n,m_1^2) T(n,m_3^2) \nonumber \\
 &&{\kern-10pt}+ (n-3) m_3 (2m_3 -m_1-m_2) \G{3}
  - \frac{(n-2)^2}{4m_1m_2} T(n,m_1^2) T(n,m_2^2) \ ,
\labbel{5} \end{eqnarray}
\begin{eqnarray}
 && \frac{m_3-m_1-m_2}{2} \frac{\partial}{\partial m_3} \G{0} =
          (n-3) \G{0} \nonumber \\
  &&{\kern-30pt} + m_1^2 \G{1} + m_2^2 \G{2} + m_3(m_1+m_2)\G{3}
\labbel{5a} \end{eqnarray}
and, for \( i=1,2,3 \),
\begin{eqnarray}
 && {\kern-10pt}
 (m_3-m_1-m_2)(m_3-m_1)(m_3-m_2)\frac{\partial}{\partial m_3} \G{i} =
        \nonumber \\
 && {\kern-20pt}
 P_{i,1}(n,m_1,m_2,m_3) \G{1} + P_{i,2}(n,m_1,m_2,m_3) \G{2} \nonumber \\
 && {\kern-20pt} +P_{i,3}(n,m_1,m_2,m_3) \G{3}
    +Q_{i,1}(n,m_1,m_2,m_3) T(n,m_2^2) T(n,m_3^2) \nonumber \\
 && {\kern-20pt} +Q_{i,2}(n,m_1,m_2,m_3) T(n,m_1^2) T(n,m_3^2)
     +Q_{i,3}(n,m_1,m_2,m_3) T(n,m_1^2) T(n,m_2^2) \ ,
\labbel{2} \end{eqnarray}
where \( T(n,m^2) \) is defined in \Eq{Tn}, while
the rational functions
\(P_{i,j}(n,m_1,m_2,m_3) \) and \(Q_{i,j}(n,m_1,m_2,m_3) \)
(which differ from the functions of the same name, but different arguments,
appearing in \cita{CCLR}) are defined as

\begin{eqnarray}
  &&P_{1,1}(n,m_1,m_2,m_3) = -\frac{n-3}{2}m_2(m_3-m_2)-(m_3-m_1)(m_3-m_2)
   \nonumber \\
  &&P_{1,2}(n,m_1,m_2,m_3) = \frac{n-3}{2}m_2(m_3-m_1)
   \nonumber \\
  &&P_{1,3}(n,m_1,m_2,m_3) = -(n-3)\left( m_3(m_3-m_1-m_2)
      + \frac{1}{2}m_2(m_1+m_2) \right)
   \nonumber \\
  &&P_{2,1}(n,m_1,m_2,m_3) = \frac{n-3}{2}m_1(m_3-m_2)
   \nonumber \\
  &&P_{2,2}(n,m_1,m_2,m_3) = -\frac{n-3}{2}m_1(m_3-m_1)-(m_3-m_1)(m_3-m_2)
   \nonumber \\
  &&P_{2,3}(n,m_1,m_2,m_3) = -(n-3)\left( m_3(m_3-m_1-m_2)
      + \frac{1}{2}m_1(m_1+m_2) \right)
   \nonumber \\
  &&P_{3,1}(n,m_1,m_2,m_3) = -(n-3)\frac{m_1}{m_3}
     (m_3-m_2)\left(m_3-\frac{1}{2}m_1-m_2\right)
   \nonumber \\
  &&P_{3,2}(n,m_1,m_2,m_3) = -(n-3)\frac{m_2}{m_3}
     (m_3-m_1)\left(m_3-m_1-\frac{1}{2}m_2\right)
   \nonumber \\
  &&P_{3,3}(n,m_1,m_2,m_3) = \frac{1}{m_3} (m_3-m_1)(m_3-m_2)(m_1+m_2-2m_3)
   \nonumber \\
  &&+\frac{n-3}{2 m_3}
     \left( 4(m_3-m_1)(m_3-m_2)(m_3-m_1-m_2)+m_1 m_2 (m_1+m_2-2m_3)
  \right)
    \ ,
 \labbel{3} \end{eqnarray}
and
\begin{eqnarray}
  &&Q_{1,1}(n,m_1,m_2,m_3) = -\frac{(n-2)^2}{8m_2m_3^2}(m_3-m_2) \nonumber \\
  &&Q_{1,2}(n,m_1,m_2,m_3) = \frac{(n-2)^2}{4m_1^2m_3^2}
        (m_3-m_1)(m_3-\frac{1}{2}m_2) \nonumber \\
  &&Q_{1,3}(n,m_1,m_2,m_3) = -\frac{(n-2)^2}{8m_1^2m_2}(m_2-m_1) \nonumber \\
  &&Q_{2,1}(n,m_1,m_2,m_3) = \frac{(n-2)^2}{4m_2^2m_3^2}
        (m_3-m_2)(m_3-\frac{1}{2}m_1) \nonumber \\
  &&Q_{2,2}(n,m_1,m_2,m_3) = -\frac{(n-2)^2}{8m_1m_3^2}(m_3-m_1) \nonumber \\
  &&Q_{2,3}(n,m_1,m_2,m_3) = -\frac{(n-2)^2}{8m_1m_2^2}(m_1-m_2) \nonumber \\
  &&Q_{3,1}(n,m_1,m_2,m_3) = \frac{(n-2)^2}{4m_2m_3^3}
       (m_3-m_2)(m_3-\frac{1}{2}m_1) \nonumber \\
  &&Q_{3,2}(n,m_1,m_2,m_3) = \frac{(n-2)^2}{4m_1m_3^3}
       (m_3-m_1)(m_3-\frac{1}{2}m_2) \nonumber \\
  &&Q_{3,3}(n,m_1,m_2,m_3) = -\frac{(n-2)^2}{4m_1m_2m_3}
       (m_3-\frac{1}{2}m_1-\frac{1}{2}m_2) \ .
 \labbel{4} \end{eqnarray}

The three \Eq{2} form a system of equations involving only the
master amplitudes \( \G{i} \quad i=1,2,3 \) and not \( \G{0} \),
while \Eq{5} can be regarded as an identity expressing \( \G{0} \) in
terms of the \( \G{i} \); that means that only three of the four master
amplitudes are independent at the pseudothreshold and, as a consequence,
\Eq{5a} is implied by the other four equations and is not an independent
equation.                                                          \par
As any system of \( k \) linear first order equations is equivalent to a
single \( k \)-th order equation, we transform the system \Eq{2}
into a single third order differential equation for a single unknown
function, which we choose to be \( \G{0} \); further, we use \Eq{5} and
\Eq{5a} for expressing the \( \G{i} \) in terms of \( \G{0} \) and
its derivatives with respect to \( m_3 \). Once the equation for
\( \G{0} \), is solved, {\it i.e.} \( \G{0} \) is known explicitly, the
\( \G{i} \) can be easily recovered, as we show later.
\par
\newcommand{\GPI}[2]{ G^{#1}_{#2}(m_1,m_2,m_3)}
Before continuing, we expand all the functions in \( n \)
around \( n=4 \); according to the notation of \cita{CCLR} we write
\begin{eqnarray}
 \G{\alpha} &=& C^2(n)\Biggl[
  \frac{1}{(n-4)^2}\GPI{(-2)}{\alpha} + \frac{1}{(n-4)}\GPI{(-1)}{\alpha}
   \nonumber \\
   &&+ \GPI{(0)}{\alpha} + O(n-4)\Biggr] \ \ \  \alpha=0,1,2,3 \ .
 \labbel{6} \end{eqnarray}
The function \(C(n)\) has been already introduced above; again from
\cita{CCLR}, where the singular parts of \( \F{\alpha} \) are given for
arbitrary values of \( p^2\), we have
\begin{eqnarray}
   &&{\kern-20pt}\GP{(-2)} = -\frac{1}{8}(m_1^2+m_2^2+m_3^2)  \ , \nonumber \\
  &&{\kern-20pt}\GP{(-1)} = -\frac{1}{32}(m_1+m_2-m_3)^2
           + \frac{3}{16}(m_1^2+m_2^2+m_3^2) \nonumber \\
    &&{\kern+90pt} -\frac{1}{8}\left[
      m_1^2\log(m_1^2)+m_2^2\log(m_2^2)+m_3^2\log(m_3^2) \right] \ ,
 \nonumber \\
 &&{\kern-20pt}\GPI{(-2)}{i} = \frac{1}{8} \ \ ,
 \ \ \GPI{(-1)}{i} = -\frac{1}{16}
  +\frac{1}{8}\log(m_i^2) \ \ , \ \ i=1,2,3.
 \labbel{7} \end{eqnarray}

The third order differential equation for \( \GP{(0)} \) then reads
\begin{eqnarray}
 &&{\kern10pt}g_3(m_1,m_2,m_3) \frac{\partial^3}{\partial m_3^3} \GP{(0)}
 + g_2(m_1,m_2,m_3) \frac{\partial^2}{\partial m_3^2} \GP{(0)} \nonumber \\
 && + g_1(m_1,m_2,m_3) \frac{\partial}{\partial m_3} \GP{(0)}
  + g_0(m_1,m_2,m_3)  \GP{(0)} \nonumber\\
 && + g(m_1,m_2,m_3) = 0 \ ,
\labbel{8} \end{eqnarray}
were the \(g_\alpha(m_1,m_2,m_3) \quad \alpha=0,..,3\) are the following
polynomials
\begin{eqnarray}
  g_3(m_1,m_2,m_3){\kern-10pt} &=&{\kern-10pt} m_3(m_3-m_1)(m_3-m_2)x^3
 \nonumber \\
  g_2(m_1,m_2,m_3){\kern-10pt} &=& {\kern-10pt}(m_3-m_1)(m_3-m_2)\left[5m_3+\frac{1}{2}(m_1+m_2)\right]
   x^2   \nonumber \\
  g_1(m_1,m_2,m_3){\kern-10pt} &=&{\kern-10pt} -2 x^4 +  2(m_1+m_2)x^3
   +  (3m_1^2+8m_1m_2+3m_2^2)x^2
   +  4 m_1 m_2(m_1+m_2)x  \nonumber \\
  g_0(m_1,m_2,m_3){\kern-10pt} &=&{\kern-10pt} -6\left(m_3-\frac{1}{2}(m_1+m_2)\right)
   \left(m_3^2-m_3(m_1+m_2)+\frac{1}{3}m_1m_2\right) \ ,
 \labbel{9} \end{eqnarray}
with \( x = m_3 - m_1 - m_2 \), while the function \( g(m_1,m_2,m_3) \),
which is written in the Appendix, is a combination of similar
polynomials and of first and second powers of the logarithms of the masses
generated in the expansion in \( (n-4) \) of the \( T(n,m^2) \).
\newcommand{\aaa}{ a(m_1,m_2,m_3)}
In order to solve \Eq{8}, we make the ansatz
\begin{equation}
   \GP{(0)} = \aaa \GP{(0,1)}
\labbel{10.1}\end{equation}
with the idea of fixing the otherwise undetermined function
\( \aaa \) by imposing that in the resulting differential equation
for \( \GP{(0,1)} \) the coefficient of the term without derivatives
vanishes, {\it i.e.} by imposing
\begin{eqnarray}
 &&{\kern-20pt}g_3(m_1,m_2,m_3) \frac{\partial^3}{\partial m_3^3} \aaa
 + g_2(m_1,m_2,m_3) \frac{\partial^2}{\partial m_3^2} \aaa \nonumber \\
 &&{\kern-20pt}+ g_1(m_1,m_2,m_3) \frac{\partial}{\partial m_3} \aaa
 + g_0(m_1,m_2,m_3)  \aaa = 0 \ ,
\labbel{10.2} \end{eqnarray}
which of course is nothing but the homogeneous version of \Eq{8}. We further
look for \( \aaa \) in the form of a rational function in \( m_3 \) and
\( m_1,m_2 \); by trial and error we find that \Eq{10.1} can be given the
explicit form
\begin{equation}
  \GP{(0)} = -\frac{2m_3-m_1-m_2}{2(m_3-m_1-m_2)^2} \GP{(0,1)} .
\labbel{10.3}\end{equation}
With the above substitution \Eq{8} has taken the form of a second order
differential equation in the first \( m_3 \)-derivative (the essential
property of the transformation \Eq{10.3} is that
\( \GP{(0,1)} \) does not appear anymore in the equation).
By iteration of the method, the following chain of substitutions is obtained
\begin{eqnarray}
 \frac{\partial}{\partial m_3} \GP{(0,1)}  &=& \GP{(0,2)}\nonumber \\
   \GP{(0,2)} &=& \left[ \frac{1}{4}(2m_3-m_1-m_2)^2 + a_0
  + \frac{a_1}{(2m_3-m_1-m_2)^2} \right] \nonumber \\
    && \GP{(0,3)} \nonumber \\
   \frac{\partial}{\partial m_3} \GP{(0,3)}&=& \GP{(0,4)}    \nonumber \\
    \GP{(0,4)} &=& \frac{1}{2} a_2 \GP{(0,5)}
\labbel{11} \end{eqnarray}
where
\begin{eqnarray}
 a_0 &=& -\frac{1}{6}(3m_1^2+2m_1m_2  + 3m_2^2) \nonumber \\
 a_1 &=& \frac{1}{12}(m_1+3m_2)(3m_1+m_2)(m_1-m_2)^2
 \nonumber \\
 a_2 &=& \frac{(2m_3-m_1-m_2)\sqrt{m_3(m_3-m_1-m_2)}}
  { (m_3-m_1)^2(m_3-m_2)^2 [m_3^2-(m_1+m_2)m_3 -\frac{1}{3}m_1m_2]^2 } \ ;
\labbel{12} \end{eqnarray}
\(\GP{(0,5)}\), finally, satisfies the following first order
differential equation
\begin{eqnarray}
 \frac{\partial}{\partial m_3} \GP{(0,5)} &=&
  (m_3(m_3-m_1-m_2))^{-\frac{3}{2}}\Bigl[ q_0 + q_1\log(m_3)
    +q_2 \log^2(m_3)\Bigr] \nonumber \\
 &&+(m_3(m_3-m_1-m_2))^{-\frac{1}{2}}\Bigl[q_{-1,0}m_3^{-1}+
  q_{-1,1}m_3^{-1}\log(m_3) \nonumber \\
 &&+ q_{-1,2}m_3^{-1}\log^2(m_3)
  +q_{0,0} + q_{0,1}\log(m_3) + q_{0,2}\log^2(m_3) \nonumber \\
  &&+q_{1,0}m_3 + q_{1,1}m_3\log(m_3) + q_{1,2}m_3\log^2(m_3) \nonumber \\
  &&+q_{2,0}m_3^2 + q_{2,1}m_3^2\log(m_3) + q_{2,2}m_3^2\log^2(m_3)
   \nonumber \\
  &&+q_{3,0}m_3^3 + q_{3,1}m_3^3\log(m_3) + q_{3,2}m_3^3\log^2(m_3)
  \labbel{13} \\
  &&+q_{4,0}m_3^4 + q_{4,1}m_3^4\log(m_3)
  +q_{5,0}m_3^5 + q_{5,1}m_3^5\log(m_3)  \Bigr]  \nonumber
\end{eqnarray}
where  the coefficients \( q_{\alpha},\ q_{\alpha , \beta}\) are functions
 of \(m_1\) and \(m_2\) only and are reported in the Appendix.

\Eq{13} is lengthy, but in fact it is a very simple differential equation,
which can immediately be solved by quadrature.
From Eq.(\ref{11}) and Eq.(\ref{13}) is obvious that we need to integrate
three times in \( m_3 \) to get \(\GP{(0)}\). The integrations are
elementary and the integration constants can be fixed at \( m_3 = m_1+m_2 \);
indeed, we are considering the pseudothreshold value
\( p^2 = -(m_1+m_2-m_3)^2 \), which for that value of \( m_3 \) corresponds
to \( p^2=0 \), where the master integrals of the sunrise graph are
known in closed analytic form \cita{CCLR}.
Once the integration constants are fixed,
the explicit form of \( \GP{(0)}\) reads
\begin{eqnarray}
 &&\GP{(0)} = \nonumber \\
   &&-\frac{1}{8(m_3-m_1-m_2)^2}
   \left[ m_1^3(m_1+2m_2){\cal L}(m_1,m_2,m_3)
         +m_2^3(2m_1+m_2){\cal L}(m_2,m_1,m_3) \right] \nonumber \\
   &&-\frac{1}{4(m_3-m_1-m_2)}
   \Bigl[ (m_1+m_2)\left(m_1^2{\cal L}(m_1,m_2,m_3)
         +m_2^2{\cal L}(m_2,m_1,m_3)\right) \nonumber \\
   &&+\frac{1}{2}(m_1^2+m_2^2+m_1m_2)\left(m_1\log\left(\frac{m_3}{m_1}\right)
     +m_2\log\left(\frac{m_3}{m_2}\right) \right)        \Bigr]  \nonumber \\
 &&+\frac{1}{8}(m_3^2-m_1^2-m_2^2)\left({\cal L}(m_1,m_2,m_3)
         + {\cal L}(m_2,m_1,m_3)\right) \nonumber \\
  &&-\frac{1}{32}\Bigl[ m_1^2\log^2(m_1^2)+ m_2^2\log^2(m_2^2)
  + m_3^2\log^2(m_3^2)
 +(m_1^2+m_2^2-m_3^2)\log(m_1^2)\log(m_2^2) \nonumber \\
 &&+(m_1^2-m_2^2+m_3^2)\log(m_1^2)\log(m_3^2)
 +(-m_1^2+m_2^2+m_3^2)\log(m_2^2)\log(m_3^2) \nonumber \\
 &&-m_1(7m_1+2m_3)\log(m_1^2)
 -m_2(7m_2+2m_3)\log(m_2^2) \nonumber \\
 &&+(2m_1^2 + 2m_1m_2 + 2m_2^2 -5m_3^2)\log(m_3^2)\Bigr]\nonumber \\
  &&-\frac{11}{128}(m_1^2+m_2^2+m_3^2) -\frac{13}{64}(-m_1m_2+m_1m_3+m_2m_3)
\labbel{15} \end{eqnarray}
where
\begin{eqnarray}
 {\cal L}(m_1,m_2,m_3) &=& \hbox{Li}_2\left(1-\frac{m_3}{m_2}\right)
                          -\hbox{Li}_2\left(-\frac{m_1}{m_2}\right)
             + \log\left(\frac{m_3}{m_1+m_2}\right)
                \log\left(\frac{m_1}{m_2}\right)
\labbel{16} \end{eqnarray}
The above result is in agreement with \cite{BDU}. To check it we have
simply to expand in \( n \) around \(n=4\) our overall coefficient \( C(n) \)
( Eq.(9) of \cita{CCLR})
and the \( n \)-dependent overall coefficients of \cite{BDU}).
\par
As already mentioned, once \(\G{0}\) is known in closed analytic form,
its derivatives are also known, and the other
amplitudes \( \G{1}\), \( \G{2}\) and \( \G{3}\) can be easily obtained
through simple relations with them.

To obtain such relations,
we differentiate \(\G{0}\) in its integral representation with respect to
\(m_1\) and \(m_2\), as done already in Eq.(\ref{5a}) for \( m_3\),
and we get after some algebra the following two additional relations

\begin{eqnarray}
 &&{\kern-20pt}\frac{\partial}{\partial m_1} \G{0} =
  \frac{-2}{m_3-m_1-m_2}
  \Biggl[ (n-3) \G{0} \nonumber \\
  &&{\kern-20pt}+ m_1(m_3-m_2) \G{1}
  + m_2^2 \G{2} +m_3^2\G{3} \Biggr]  \nonumber \\
 &&{\kern-20pt}\frac{\partial}{\partial m_2} \G{0} =
  \frac{-2}{m_3-m_1-m_2}
  \Biggl[ (n-3) \G{0} \nonumber \\
  &&{\kern-20pt}+ m_1^2 \G{1}
   + m_2(m_3-m_1) \G{2} +m_3^2\G{3} \Biggr] \ .
\labbel{17} \end{eqnarray}

 The first of the \Eq{2}, \Eq{5} and \Eq{17} form a system of equations,
 which allows to express master amplitudes \( \G{\alpha}, \alpha=0,1,2,3\),
 as a function of first derivatives of \( \G{0} \) only.
 As \(\G{0}\) is already known, we can express \( \G{i}, i=1,2,3\),
 as a function of \(\G{0}\) and two of its first derivatives (we choose
 derivatives in respect to \( m_1\) and \(m_2\) ). The solution reads
 (the expression for \(\G{2} \) can be easily found from \(\G{1}\) by
  the substitution \( m_1 \rightarrow m_2 \ \ , \ m_2 \rightarrow m_1 \) )

\begin{eqnarray}
 &&{\kern-18pt}\G{1} = \nonumber \\
 &&{\kern-18pt}\frac{1}{8m_1m_3}\Biggl\{
     \left(m_1-3m_3\right)\frac{\partial}{\partial m_1} \G{0}
  +\left(m_2+m_3\right)\frac{\partial}{\partial m_2} \G{0}
   \nonumber \\
 &&{\kern-21pt}+\frac{1}{m_3-m_1-m_2}\Biggl[\Bigl(
                                   2(n-3)(m_1+m_2)-(n-4)m_3\Bigr)
    \G{0}  \nonumber \\
 &&{\kern-21pt}+\frac{(n-2)^2}{2(n-3)}\Biggl(\frac{1}{m_2}T(n,m_2^2)T(n,m_3^2)
  +\frac{1}{m_1}T(n,m_1^2)T(n,m_3^2)
  -\frac{m_3}{m_1m_2}T(n,m_1^2)T(n,m_2^2)
   \Biggr) \Biggr] \Biggr\} \nonumber \\
 &&{\kern-18pt}\G{3} = \nonumber \\
 &&{\kern-18pt}\frac{1}{8m_3^2}\Biggl\{
     \left(3m_1-m_3\right)\frac{\partial}{\partial m_1} \G{0}
   +\left(3m_2-m_3\right)\frac{\partial}{\partial m_2} \G{0}
   \nonumber \\
 &&{\kern-21pt}+\frac{1}{m_3-m_1-m_2}\Biggl[\Bigl(
                                  (n-3)(6m_1+6m_2-7m_3)-m_3\Bigr)
    \G{0} \labbel{18} \\
 &&{\kern-21pt}-\frac{(n-2)^2}{2(n-3)}\Biggl(\frac{1}{m_2}T(n,m_2^2)T(n,m_3^2)
  +\frac{1}{m_1}T(n,m_1^2)T(n,m_3^2)
  -\frac{m_3}{m_1m_2}T(n,m_1^2)T(n,m_2^2)
   \Biggr) \Biggr] \Biggr\} \nonumber
  \end{eqnarray}
Expanding around \(n=4\) we find that \(\GPI{(-2)}{\alpha} \) and
\noindent
 \(\GPI{(-1)}{\alpha} \), (\( \alpha=0,1,2,3\)), already explicitly given in
\Eq{7}, automatically satisfy the \Eq{18}, while for
\noindent
\(\GPI{(0)}{i} \), (\( i=1,2,3\)), we find the explicit expressions
\begin{eqnarray}
 && {\kern-20pt}G_1^{(0)}(m_1,m_2,m_3) = \nonumber \\
   &&\frac{1}{8(m_3-m_1-m_2)^2}
   \left[ m_1(m_1+2m_2){\cal L}(m_1,m_2,m_3)
         -m_2^2{\cal L}(m_2,m_1,m_3) \right] \nonumber \\
   &&+\frac{1}{4(m_3-m_1-m_2)}
   \Biggl[ (m_1+m_2){\cal L}(m_1,m_2,m_3)
   +\frac{1}{2}\left( m_1\log\left(\frac{m_3}{m_1}\right)
     +m_2\log\left(\frac{m_3}{m_2}\right) \right)        \Biggr]  \nonumber \\
 &&+\frac{1}{8}\Bigl({\cal L}(m_1,m_2,m_3)
         + {\cal L}(m_2,m_1,m_3)\Bigr)
  +\frac{1}{32}\Bigl[ \log^2(m_1^2)
 + \log(m_1^2)\log(m_2^2) \nonumber \\
 &&+\log(m_1^2)\log(m_3^2)
  -\log(m_2^2)\log(m_3^2)
  -4\log(m_1^2)
  +2\log(m_3^2)-1\Bigr] \nonumber \\
 &&{\kern-20pt} G_3^{(0)}(m_1,m_2,m_3) = \nonumber \\
   &&\frac{1}{8(m_3-m_1-m_2)^2}
   \left[ m_1^2{\cal L}(m_1,m_2,m_3)
         +m_2^2{\cal L}(m_2,m_1,m_3) \right] \nonumber \\
   &&+\frac{1}{8(m_3-m_1-m_2)}
    \left( m_1\log\left(\frac{m_3}{m_1}\right)
     +m_2\log\left(\frac{m_3}{m_2}\right) \right)   \nonumber \\
 &&-\frac{1}{8}\Bigl({\cal L}(m_1,m_2,m_3)
         + {\cal L}(m_2,m_1,m_3)\Bigr)
  +\frac{1}{32}\Bigl[ \log^2(m_3^2)
  -\log(m_1^2)\log(m_2^2) \nonumber \\
  &&+\log(m_1^2)\log(m_3^2)
  +\log(m_2^2)\log(m_3^2)
  -2\log(m_3^2)-1\Bigr]  \ ,
\labbel{19} \end{eqnarray}

where \({\cal L}(m_1,m_2,m_3)\) is defined by \Eq{16}. The function
\( G_2^{(0)}(m_1,m_2,m_3)\) is obtained by substituting
\( m_1 \rightarrow m_2 \ \ , \ m_2 \rightarrow m_1 \) in the expression of
\( G_1^{(0)}(m_1,m_2,m_3)\).
\section {The expansion at the pseudothreshold.}
As recalled in \cita{CCLR}, it is easy to find the coefficients of the
expansion of a function in any variable once a system of differential
equations in that variable is given. In this section we
obtain the expansions in \( p^2 \) at the pseudothreshold
of all the master integrals
\Eq{Falpha}, \Eq{expa}. Putting expansions \Eq{expa} into the system
of equations \Eq{Feqs} we get (comparing the lowest terms in expansion)
from the first equation
\begin{eqnarray}
 \HH{0,1} &=& -\frac{1}{(m_3-m_1-m_2)^2} \Bigl[
 \left(n-3\right) \GG{0} \nonumber \\
 &&{\kern-120pt}+ m_1^2 \GG{1} + m_2^2 \GG{2} + m_3^2 \GG{3}\Bigr]
   \ .
\labbel{EPS1} \end{eqnarray}
Substituting for the \( \GG{\alpha} \) their expressions expanded
around \(n=4\), we have
\begin{equation}
 \HH{0,1} = C^2(n)\Biggl[ \frac{1}{32 \ (n-4)}
  +H^{(0,1)}(n=4,m_1,m_2,m_3) + {\cal O}(n-4) \Biggr] \ ,
\labbel{PS1a} \end{equation}
with
\begin{eqnarray}
 H^{(0,1)}(n=4,m_1,m_2,m_3) &=& \nonumber \\
 &&{\kern-140pt}-\frac{1}{8(m_3-m_1-m_2)^4}
   \left[ m_1^3(m_1+2m_2){\cal L}(m_1,m_2,m_3)
         +m_2^3(2m_1+m_2){\cal L}(m_2,m_1,m_3) \right] \nonumber \\
 &&{\kern-140pt}+\frac{1}{8(m_3-m_1-m_2)^3}\Biggl[
  -2(m_1+m_2)\left(m_1^2{\cal L}(m_1,m_2,m_3)+m_2^2{\cal L}(m_2,m_1,m_3)\right)
 \nonumber \\
 &&{\kern-40pt}+ m_1 m_2(m_1+m_2) \log\left(\frac{m_1 m_2}{m_3^2}\right)
   + m_1^3 \log\left(\frac{m_1 }{m_3}\right)
            + m_2^3 \log\left(\frac{m_2 }{m_3}\right) \Biggr]
 \nonumber \\
 &&{\kern-140pt}-\frac{1}{8(m_3-m_1-m_2)^2}\Biggl[
  m_1^2{\cal L}(m_1,m_2,m_3)+m_2^2{\cal L}(m_2,m_1,m_3)
 -\frac{1}{2}\left(m_1^2+m_1 m_2+m_2^2\right)
 \nonumber \\
   &&{\kern-40pt}
 - m_1 \left(\frac{3}{2}m_1+m_2\right) \log\left(\frac{m_1 }{m_3}\right)
            - m_2 \left(m_1+\frac{3}{2}m_2\right)
    \log\left(\frac{m_2 }{m_3}\right) \Biggr]
  \nonumber \\
 &&{\kern-140pt}+\frac{m_1+m_2}{16(m_3-m_1-m_2)}
  \ \ +  \frac{1}{16}\log(m_3)  -  \frac{5}{128}
   \ .
\labbel{PS1b} \end{eqnarray}
The other three equations form a system of linear equations,
which can be easily solved algebraically.
The solution for \(\HH{1,1}\) reads
\begin{eqnarray}
 \HH{1,1} &=& -\frac{1}{m_1 (m_3-m_1-m_2)^2} \Biggl[ \nonumber \\
&&{\kern-80pt} \frac{1}{n-4}\Biggl( \frac{1}{16}\left(m_3 - 2 m_1 - m_2\right)
  \GG{1}\nonumber \\
 &&{\kern-55pt}+ \frac{1}{16}\left(m_3 - m_1 - 2 m_2\right)  \GG{2}\nonumber \\
 &&{\kern-55pt}+ \frac{1}{16}\left(2 m_3 - m_1 - m_2\right)  \GG{3}\Biggr)
  + \cdots \Biggr] \ .
\labbel{EPS2} \end{eqnarray}
We do not put here the rest of the expression as it is not relevant for
the subsequent discussion. Similar formulae are found for \(\HH{2,1}\)
and \(\HH{3,1}\).

As it is evident from the \Eq{EPS2}, to get
terms \( \sim (n-4)^0 \) in the \( (n-4) \) expansion of \( \HH{1,1} \)
we need terms \( \sim (n-4) \) of the \( \GG{i}\ , i = 1,2,3\),
(a similar problem was found in \cita{DS} for the expansion
at the threshold).  The terms \( \sim (n-4) \) are not yet known
and could in principle be investigated with the method used in the
previous section. The integrals obtained that way
contain up to trilogarithmic functions (which are however expected to
cancel out in the combination appearing in the {\it r.h.s.} of
\Eq{EPS2} - see below), so we decided to get
\( \HH{i,1} \ , i = 1,2,3 \), in a simpler way.
By using \Eq{Fi} we get the following relations
 \begin{eqnarray}
 \HH{1,1} &{\kern-10pt}=&{\kern-14pt}
   - \frac{\partial \HH{0,1}}{\partial m_1^2}
   + 2\frac{m_3{\kern-5pt}-{\kern-4pt}m_1{\kern-5pt}-{\kern-4pt}m_2}
   {m_1} \HH{0,2} \nonumber \\
 \HH{2,1} &{\kern-10pt}=&{\kern-14pt}
  - \frac{\partial \HH{0,1}}{\partial m_2^2}
   + 2\frac{m_3{\kern-5pt}-{\kern-4pt}m_1{\kern-5pt}
  -{\kern-4pt}m_2}{m_2} \HH{0,2} \nonumber \\
 \HH{3,1} &{\kern-10pt}=&{\kern-14pt}
  - \frac{\partial \HH{0,1}}{\partial m_3^2}
   - 2\frac{m_3{\kern-5pt}-{\kern-4pt}m_1{\kern-5pt}
  -{\kern-4pt}m_2}{m_3} \HH{0,2} \ . \nonumber \\
 &&\phantom{}
 \labbel{EPS3}  \end{eqnarray}
As the pole terms in \( (n-4) \) expansion of
\( \HH{i,1} \ , i = 1,2,3 \), are known to
vanish \cita{CCLR}, from \Eq{EPS3} and with the result in
\Eq{PS1a} we deduce that \( \HH{0,2} \) has no pole at \(n=4\);
the \Eq{EPS3} takes the form for \(n=4\)
\begin{eqnarray}
H^{(1,1)}(n=4,m_1,m_2,m_3) &=&  C^2(n)\Biggl[
 - \frac{\partial H^{(0,1)}(n=4,m_1,m_2,m_3)}{\partial m_1^2}
\nonumber \\
   && + 2\frac{m_3 -m_1 -m_2}{m_1} H^{(0,2)}(n=4,m_1,m_2,m_3) \Biggr]
\nonumber \\
H^{(2,1)}(n=4,m_1,m_2,m_3) &=&  C^2(n)\Biggl[
 - \frac{\partial H^{(0,1)}(n=4,m_1,m_2,m_3)}{\partial m_2^2}
\nonumber \\
   && + 2\frac{m_3 -m_1 -m_2}{m_2} H^{(0,2)}(n=4,m_1,m_2,m_3) \Biggr]
\nonumber \\
H^{(3,1)}(n=4,m_1,m_2,m_3) &=&  C^2(n)\Biggl[
 - \frac{\partial H^{(0,1)}(n=4,m_1,m_2,m_3)}{\partial m_3^2}
\nonumber \\
   && - 2\frac{m_3 -m_1 -m_2}{m_3} H^{(0,2)}(n=4,m_1,m_2,m_3) \Biggr] \ .
 \labbel{EPS3a}  \end{eqnarray}
Therefore to find  \( H^{(i,1)}(n=4,m_1,m_2,m_3) \), \( i = 1,2,3 \),
it is enough to insert the derivatives in respect to the masses,
easily obtained from \Eq{PS1b}, and to have the explicit form of
\( H^{(0,2)}(n=4,m_1,m_2,m_3) \).
We evaluate it in the next section using dispersion relation method.

There is no further obstacle in obtaining higher order terms in the \( p^2 \)
expansion at the pseudothreshold,
as can be verified with an explicit algebraic calculation,
by solving iteratively
the system of linear equations, obtaining the coefficients of
any order in terms of the coefficients of lower orders.
The reason is that the troublesome denominator \( 1/(n-4) \)
appears only when expressing the first order coefficients in terms of the
zeroth order coefficients; the denominators appearing at higher orders
are \( 1/(n-5), 1/(n-6), \cdots \), regular at \( n=4 \).

The explicit expressions of the solutions are easily obtained with an
algebraic program, but they are somewhat lengthy and we do not present
them here for the sake of brevity. Let us also observe that
for the practical purposes of solving the system of differential equations
numerically
only the values at pseudothresholds and first derivatives are needed.

It is worth to note that the functions \( G_{i}^{0}(m_1,m_2,m_3) \ ,
i=1,2,3 \) are not independent (the pole terms in \Eq{EPS2}
have to cancel) and the relation between them, using \Eq{6} and \Eq{19},
reads

 \begin{eqnarray}
      0=   &&\GPI{(0)}{1}  (  - 8 m_1 - 4 m_2 + 4 m_3 ) \nonumber \\
        + &&\GPI{(0)}{2} (  - 4 m_1 - 8 m_2 + 4 m_3 ) \nonumber \\
       + &&\GPI{(0)}{3}  (  - 4 m_1 - 4 m_2 + 8 m_3 ) \nonumber \\
       + &&\frac{1}{2} \Bigl[ (  - m_1 - m_2 + m_3 ) \nonumber \\
       &&{\kern+30pt}+ \log(m_1^2) (  - m_2 + m_3 )
       + \log(m_2^2) (  - m_1 + m_3 )
       - \log(m_3^2) (  m_1 + m_2 ) \Bigr] \nonumber \\
       + &&\frac{1}{8} \Bigl[ \log^2(m_1^2) ( 2 m_1 + m_2 - m_3 ) \nonumber \\
                            &&{\kern+30pt}+ \log^2(m_2^2) ( m_1 + 2 m_2 - m_3 )
                            + \log^2(m_3^2) ( m_1 + m_2 - 2 m_3 )
      \Bigr]   \nonumber \\
      + &&\frac{1}{4} \Bigl[ \log(m_1^2) \log(m_2^2)  ( m_1 + m_2 )
             \nonumber \\
                          &&{\kern+30pt} + \log(m_1^2) \log(m_3^2)  ( m_1 - m_3 )
                           + \log(m_2^2) \log(m_3^2)  ( m_2 - m_3 )
   \Bigr] \ .
 \labbel{EPS4}  \end{eqnarray}

 This relation is of course fulfilled by the already known solutions
 \Eq{19}.
\section{The value of  \boldmath $  H^{(0,2)}(n=4,m_1,m_2,m_3)$ .}

 Let us define the function \( \tilde F (p^2)\) through the equation

 \begin{eqnarray}
  F_{0}(n=4,m_1^2,m_2^2,m_3^2,p^2) =
  &&F_{0}(n=4,m_1^2,m_2^2,m_3^2,0) \nonumber \\
 + &p^2 &F'_{0}(n=4,m_1^2,m_2^2,m_3^2,0) \nonumber \\
 + &&\tilde F (p^2) \ .
 \labbel{V1} \end{eqnarray}

 The subtracted dispersion relation for \( \tilde F (p^2) \)
 reads

 \begin{equation}
 \tilde F (p^2) = (p^2)^2 \ \ \ \int\limits_{(m_1+m_2+m_3)^2}^{\infty}
  \frac{du}{u^2 (u+p^2)} \ \
  \frac{1}{\pi} \ \  {\mathrm Im}  F_{0}(n=4,m^2_1,m^2_2,m^2_3,u)
  \ ,
 \labbel{V2} \end{equation}

 where

 \begin{equation}
      \frac{1}{\pi} \ \ {\mathrm Im}  F_{0}(n=4,m^2_1,m^2_2,m^2_3,u) =
 \frac{1}{16} \ \int\limits_{(m_1+m_2)^2}^{(\sqrt{u} -m_3)^2} db
 \ \ \frac{R(u,b,m_3^2)}{u} \ \ \frac{R(b,m_1^2,m_2^2)}{b}  \ ,
\labbel{V3} \end{equation}

with the usual definition of the function \( R\)
\begin{equation}
   R(x,y,z) = \sqrt{x^2 + y^2 + z^2 - 2xy - 2xz - 2yz} \ .
\labbel{V6} \end{equation}

The function $  H^{(0,2)}(n=4,m_1,m_2,m_3)$ which we want to calculate
in this section in the notation of \Eq{V1},\Eq{V2} reads

 \begin{eqnarray}
 &&H^{(0,2)}(n=4,m_1,m_2,m_3) = \ \ \frac{1}{2}
                              \tilde F''(p^2=-(m_3-m_1-m_2)^2) \ \ =
  \nonumber \\
  && \int\limits_{(m_1+m_2+m_3)^2}^{\infty}
  \frac{ du}{(u- (m_3-m_1-m_2)^2 )^3} \ \
  \frac{1}{\pi} \ \  {\mathrm Im}  F_{0}(n=4,m^2_1,m^2_2,m^2_3,u)
  \ ,
 \labbel{V4} \end{eqnarray}

By using \Eq{V3} and exchanging the order of integration we get
 \begin{eqnarray}
 &&H^{(0,2)}(n=4,m_1,m_2,m_3) = \nonumber \\
 &&\frac{1}{16}
\int\limits_{(m_1+m_2)^2}^{\infty} db\ \ \frac{R(b,m_1^2,m_2^2)}{b}
  \int\limits_{(\sqrt{b}+m_3)^2}^{\infty}
  du \frac{ R(u,b,m_3^2)}{u \ (u- (m_3-m_1-m_2)^2 )^3} \ \
     \ .
 \labbel{V5} \end{eqnarray}
The integration in \( u \) is elementary and gives
 \begin{equation}
 H^{(0,2)}(n=4,m_1,m_2,m_3) = \frac{1}{32}
\int\limits_{(m_1+m_2)^2}^{\infty} db\ \ \frac{R(b,m_1^2,m_2^2)}{b}
 { \tilde H(b)}
     \ ,
 \labbel{V7} \end{equation}

 where

\def\den{(m_3-m_1-m_2)}
\def\[log(yS)]{\log(y_S)}

 \begin{eqnarray}
 \tilde H(b) &=& \frac{\[log(yS)]}{ {\cal R}(b)\den^2 } \left[
         - \frac{2 ( m_3^2-b )^2 }{\den^4}
       +   \frac {2(b + m_3^2 )}{\den^2}
       +  \frac {4 b m_3^2}{{\cal R}^2(b)}  \right] \nonumber \\
       &&+ \log\left(\frac{b}{m_3^2}\right) \frac{- b + m_3^2}{\den^6}
         \nonumber \\
       &&+ \frac{2}{\den^4}
       + \frac{1}{\den^2}\frac{b + m_3^2}{{\cal R}^2(b)}
       - \frac{1}{{\cal R}^2(b)} \ ,
 \labbel{V8} \end{eqnarray}

with

 \begin{equation}
 {\cal R}(b) = R\left(b,(m_1+m_2-m_3)^2,m_3^2\right)
 \labbel{V9} \end{equation}

and
\begin{equation}
 y_S = \frac{ b + m_3^2 - (m_1+m_2-m_3)^2 -{\cal R}(b)}{ 2 \ \sqrt{b} \ m_3}
 \ .
 \labbel{V10} \end{equation}

To perform the last integration, it is convenient to differentiate
with respect to one of the masses \Eq{V7} in its integral form, to
perform the integration in \( b \) of the derivative so obtained
(the integration in  \( b \) being now elementary) and then to
integrate again in the mass. As a final result we find

\begin{eqnarray}
  &&H^{(0,2)}(n=4,m_1,m_2,m_3)=\frac{1}{32} \Biggl\{
  \frac{R^2(m_1,m_2,-m_3)}{2 \ \den^4}
  \nonumber \\
 &&+ \frac{2 \ m_1 \ \log\left(\frac{m_1}{m_3}\right)}{\den^5}
     \left( m_2^2 + m_3^2 - m_1 m_2 + m_1 m_3 \right) \nonumber \\
 &&+\frac{2 \ m_2 \ \log\left(\frac{m_2}{m_3}\right)}{\den^5}
     \left( m_1^2 + m_3^2 - m_1 m_2 + m_2 m_3 \right) \nonumber \\
  &&+ \frac{2 \ {\cal I}_0}{\den^6}
   \left( 2 m_1^2 m_2^2 - m_1^2 m_3^2 - m_2^2 m_3^2 \right)
  -\frac{2 \ m_3^2 \ {\cal I}_1}{\den^6} \left(m_1^2-m_2^2 \right) \nonumber \\
 &&-\frac{4 \ {\cal I}_2}{\den^3} \ \ \frac{m_1 m_2 m_3}
  {R\left((m_1+m_2)^2,(m_1-m_3)^2,(m_2-m_3)^2\right)} \Biggr\}
  \ ,
 \labbel{V11} \end{eqnarray}

where the integrals \( {\cal I}_i \ \ i=0,1,2 \) are defined in the
 following way

 \begin{eqnarray}
 {\cal I}_0 &=& \int\limits_{(m_1+m_2)^2}^{\infty} db\ \
 \left[ \frac{\log(y_S)} {R\left(b,(m_1-m_3)^2,(m_2-m_3)^2\right)}
       + \frac{\log\left(\frac{\sqrt{b}}{m_3}\right)}{R(b,m_1^2,m_2^2)}
 \right] \nonumber \\
 &=& \zeta_2 + \log\left(\frac{m_3}{m_1}\right)
               \log\left(\frac{m_2}{m_3}\right)
              -\hbox{Li}_2\left(1 -\frac{m_1}{m_3}\right)
              -\hbox{Li}_2\left(1 -\frac{m_2}{m_3}\right) \ ,
\labbel{V12} \end{eqnarray}

 \begin{eqnarray}
 &&{\kern-45pt}{\cal I}_1 = \int\limits_{(m_1+m_2)^2}^{\infty} \frac{db}{b}\ \
  \Biggl[ \frac{(m_1- m_2)(m_1+m_2-2 m_3)}
   {R\left(b,(m_1-m_3)^2,(m_2-m_3)^2\right)}
      \log(y_S)
       - \frac{\left(m_1^2-m_2^2\right)}{R(b,m_1^2,m_2^2)}
  \log\left(\frac{\sqrt{b}}{m_3}\right)
 \ \  \Biggr] \nonumber \\
 &&{\kern-58pt}=  -\zeta_2 +\log\left(\frac{m_1}{m_2}\right)
                   \log\left(\frac{m_2 m_3}{(m_1+m_2)^2}\right)
              -2 \hbox{Li}_2\left(-\frac{m_1}{m_2}\right)
              +\hbox{Li}_2\left(1 -\frac{m_1}{m_3}\right)
              -\hbox{Li}_2\left(1 -\frac{m_2}{m_3}\right) \ , \nonumber \\
\
 \labbel{V13} \end{eqnarray}

and

 \begin{eqnarray}
 {\cal I}_2 &=& \int\limits_{(m_1+m_2)^2}^{\infty} \frac{db}{b-(m_1+m_2)^2}\ \
 \frac{R\left((m_1+m_2)^2,(m_1-m_3)^2,(m_2-m_3)^2\right)}
      {R\left(b,(m_1-m_3)^2,(m_2-m_3)^2\right)} \ \ \log(y_S)  \nonumber \\
 &{\kern-15pt}=&{\kern-15pt}
    -5\zeta_2 + \log\left( \frac{t_2^2 \ \left(1+t_1^2\right)}
                                 {1+t_2^2}\right) \
                \log\left( \frac{1-t_1^2 \ t_2^2}{1+t_2^2}\right)
  - \log\left(\frac{1-t_1 \ t_2}{1-t_2}\right) \
       \log\left(\frac{t_2^2 \ \left(1-t_1\right)^2}
                      {\left(1-t_2\right)^2}\right)
    \nonumber \\
  &&{\kern-15pt}
 -\ 2 \  \hbox{Li}_2\left(\frac{t_2\left(1-t_1\right)}{t_2-1}\right)
     \ - 2   \ \hbox{Li}_2\left(\frac{t_1 - t_2 }{1-t_2} \right)
  +\ 2  \ \hbox{Li}_2\left(\frac{1+t_1\ t_2}{1+t_2}\right)
   \ - 2   \ \hbox{Li}_2\left(\frac{t_1+ t_2 }{1+t_2} \right)
    \nonumber \\
&&{\kern-15pt}+\ \ \hbox{Li}_2\left(\frac{t_2^2 \ \left(1+t_1^2\right)}
                                 {1+t_2^2}\right)
  +\ \ \hbox{Li}_2\left(\frac{1-t_2^2}{1+t_2^2}\right)
  +\ \ \hbox{Li}_2\left(\frac{t_2^2-t_1^2 }{1+t_2^2}\right)
  -\ \ \hbox{Li}_2\left(\frac{t_2^2-1}{1+t_2^2}\right)
    \ ,
 \labbel{V14} \end{eqnarray}

where

\begin{equation}
 t_1 = \frac{\sqrt{m_1+m_2-m_3}-\sqrt{m_3}}{\sqrt{m_1+m_2-m_3}+\sqrt{m_3}}
 \ \ \ \ \hbox{and}
 \ \ \   t_2 = \frac{\sqrt{m_1}-\sqrt{m_2}}{\sqrt{m_1}+\sqrt{m_2}}
 \ .
\labbel{V15} \end{equation}

 The last expression was found assuming \(m_1>m_2>m_3\), but the
 analytic continuation to other regions is straightforward.

\section{Summary}
 In this paper we have presented the expansion of the 2-loop sunrise selfmass
 master amplitudes at the pseudothreshold \(p^2 = -(m_1+m_2-m_3)^2\). The
 other pseudothresholds can be easily found by the permutation of the
 masses.
 We define the expansion in \Eq{expa};
 the values of the amplitudes at the pseudothreshold are given
 in \Eq{6}, \Eq{7}, \Eq{15} and \Eq{19}.
 The first order terms in the pseudothreshold expansion of the master
 amplitudes at \((n=4)\) are presented in  \Eq{PS1a}, \Eq{PS1b} and
 \Eq{EPS3a}, while the  second order term of
 \(F_{0}(n=4,m_1^2,m_2^2,m_3^2,p^2 = -(m_1+m_2-m_3)^2 )\)
 is presented in \Eq{V11}.
 The higher order terms, which are not given explicitly here, can be easily
 found by solving recursively, at each order, a system of four algebraic
 linear equations.
 The expansion at the pseudothreshold cannot be simply
 deduced from the known expansion at the threshold \cita{DS} and vice versa,
even if at first sight they seem to be connected by the change of sign
\( m_3 \to -m_3 \). In fact the analytic properties of the amplitudes
are different at the two points: at the pseudothreshold
the sunrise amplitudes are regular, so that
the solution of the \Eq{Feqs} can be expanded as a single power series,
while at the threshold the sunrise amplitudes develop a branch point
and its expansion is indeed the sum of two series, \cita{I}.

\vskip 0.4 cm

{\bf Acknowledgments.}
One of the authors (E.R.) wants to thank the Alexander von Humboldt
Foundation for supporting his stay at the Theoretische TeilchenPhysik
Institut of the Karslruhe University, where a substantial part of this
work was carried out.
H.C. is grateful for the support and the kind hospitality
to the Bologna Section of INFN and to the Department
of Physics of the Bologna University, where the work was completed.
\section{Appendix.}

In this appendix are given the definitions of the functions used
in the text.

Function \(g(m_1,m_2,m_3)\) used in \Eq{8}
\begin{eqnarray}
  g(m_1,m_2,m_3) &=& r_{0} + r_{1} \log(m_1^2) + r_{2} \log(m_2^2)
  + r_{3} \log(m_3^2) + r_{1,1}\log^2(m_1^2) \nonumber \\
  &&+ r_{1,2}\log(m_1^2) \log(m_2^2) + r_{1,3} \log(m_1^2) \log(m_3^2)
  + r_{2,2}\log^2(m_2^2) \nonumber \\
  &&+ r_{2,3}\log(m_2^2)\log(m_3^2)
  + r_{3,3}\log^2(m_3^2) \ ,
 \labbel{9a} \end{eqnarray}
where
\begin{eqnarray}
 r_{0} &=&  -\frac{11}{16} x^5 - \frac{147}{32}(m_1 + m_2)x^4
   - (\frac{485}{64}m_1^2+ \frac{211}{16}m_1m_2+ \frac{485}{64}m_2^2)x^3
  \nonumber \\
   &&- (\frac{81}{16}m_1^3+\frac{839}{64}m_1^2m_2+\frac{839}{64}m_1m_2^2
         +\frac{81}{16}m_2^3)x^2 \nonumber \\
   &&- (\frac{21}{16}m_1^4+5m_1^3m_2 +\frac{13}{2}m_1^2m_2^2
      +5m_1m_2^3 +\frac{21}{16}m_2^4)x \nonumber \\
     &&-\frac{7}{16}m_1m_2(m_1^3 +2m_1^2m_2 + 2m_1m_2^2
          +m_2^3 ) \nonumber \\
   r_{1} &=& m_1^2 \Bigl[ \frac{1}{2}x^3 + \frac{9}{8}(m_1+m_2)x^2
  +(\frac{9}{16}m_1^2 +\frac{5}{4} m_1m_2 + \frac{5}{16}m_2^2)x
  +\frac{3}{16}m_1m_2(m_1+m_2)\Bigr] \nonumber \\
   r_{2} &=& m_2^2 \Bigl[ \frac{1}{2}x^3 + \frac{9}{8}(m_1+m_2)x^2
  +(\frac{5}{16}m_1^2 +\frac{5}{4} m_1m_2 + \frac{9}{16}m_2^2)x
  +\frac{3}{16}m_1m_2(m_1+m_2)\Bigr] \nonumber \\
 r_{3} &=& \frac{15}{4}x^5 + \frac{71}{8}(m_1+m_2)x^4
   + (\frac{121}{16}m_1^2 +\frac{67}{4}m_1m_2 + \frac{121}{16}m_2^2)x^3
   \nonumber \\
 &&+(3m_1^3 +\frac{171}{16}m_1^2m_2+\frac{171}{16}m_1m_2^2 +3m_2^3)
  x^2 \nonumber \\
 &&+(\frac{9}{16}m_1^4 +\frac{11}{4}m_1^3m_2 +\frac{35}{8}m_1^2 m_2^2
     +\frac{11}{4}m_1m_2^3 +\frac{9}{16}m_2^4)x
   +\frac{3}{16}m_1m_2(m_1+m_2)^3
       \nonumber \\
 r_{1,1} &=& -\frac{3}{16}m_1^2\left(m_3-\frac{1}{2}(m_1+m_2)\right)
         \left(m_3^2-m_3(m_1+m_2)+\frac{1}{3}m_1m_2\right)
   \nonumber \\
 r_{1,2} &=& \frac{1}{8}m_1^2m_2^2\left(m_3-\frac{1}{2}(m_1+m_2)\right)
  \nonumber \\
 r_{1,3} &=& -\frac{1}{8}m_1^2\left(m_3-\frac{1}{2}(m_1+m_2)\right)
 \left(3x^2 +3(m_1+m_2)x + m_2(m_1+m_2)\right)
   \nonumber \\
 r_{2,2} &=& -\frac{3}{16}m_2^2\left(m_3-\frac{1}{2}(m_1+m_2)\right)
         \left(m_3^2-m_3(m_1+m_2)+\frac{1}{3}m_1m_2\right)
   \nonumber \\
 r_{2,3} &=& -\frac{1}{8}m_2^2\left(m_3-\frac{1}{2}(m_1+m_2)\right)
 \left(3x^2 +3(m_1+m_2)x + m_1(m_1+m_2)\right)
   \nonumber \\
 r_{3,3} &=& - \frac{1}{16}\left(m_3-\frac{1}{2}(m_1+m_2)\right) \nonumber \\
   &&\left(3(m_1^2+m_2^2)x^2 + 3(m_1^3+m_1^2m_2+m_1m_2^2+m_2^3)x
    +m_1m_2(m_1+m_2)^2\right)
   \
 \labbel{10} \end{eqnarray}
and \( x = m_3 - m_1 - m_2 \).

The definition of the coefficients used in \Eq{13}:
\begin{eqnarray}
 q_0 &=& m_1^2m_2^2(m_1+m_2)\Bigl[
        \frac{7}{48} (m_1^2 + m_1m_2 + m_2^2)
        -\frac{1}{16}m_1^2\log(m_1^2)
        -\frac{1}{16}m_2^2\log(m_2^2) \nonumber \\
         &&+\frac{1}{96}(m_1\log(m_1^2)-m_2\log(m_2^2))^2 \Bigr]
  \nonumber \\
 q_1&=& m_1^2m_2^2(m_1+m_2)^2\Bigl[
         -\frac{1}{8}(m_1+m_2)
        +\frac{1}{24}m_1 \log(m_1^2)
        +\frac{1}{24}m_2 \log(m_2^2)\Bigr] \nonumber \\
 q_2&=& \frac{1}{24}m_1^2m_2^2(m_1+m_2)^3 \nonumber \\
 q_{-1,0}&=& m_1m_2\Bigl[-\frac{1}{192}(5m_1^4 + 29m_1^3m_2 +104m_1^2m_2^2
               +29m_1m_2^3 +5m_2^4) \nonumber \\
     &&+ \frac{1}{48}m_1^2(m_1^2+5m_2^2)\log(m_1^2)
     + \frac{1}{48}m_2^2(5m_1^2+m_2^2)\log(m_2^2)  \nonumber \\
     &&+\frac{1}{48}m_1m_2(m_1\log(m_1^2)-m_2\log(m_2^2))^2 \Bigr] \nonumber \\
 q_{-1,1} &=& m_1^2m_2^2(m_1+m_2)\Bigl[ -\frac{7}{24}(m_1+m_2)
              + \frac{1}{12}\left(m_1\log(m_1^2) + m_2\log(m_2^2)\right)\Bigr]
 \nonumber \\
 q_{-1,2} &=& \frac{1}{12}m_1^2m_2^2(m_1+m_2)^2 \nonumber \\
 q_{0,0} &=& -\frac{5}{64}m_1^5 -\frac{21}{32}m_1^4m_2 -\frac{67}{48}m_1^3m_2^2
   -\frac{67}{48}m_1^2m_2^3 -\frac{21}{32}m_1m_2^4 -\frac{5}{64}m_2^5
   \nonumber \\
  &&+m_1^2 (\frac{1}{16}m_1^3 +\frac{5}{24}m_1^2m_2 +\frac{11}{24}m_1m_2^2
    +\frac{5}{16}m_2^3)\log(m_1^2) \nonumber \\
  &&+m_2^2 (\frac{5}{16}m_1^3 +\frac{11}{24}m_1^2m_2 +\frac{5}{24}m_1m_2^2
    +\frac{1}{16}m_2^3)\log(m_2^2) \nonumber \\
   &&-\frac{1}{16}m_1^2m_2^2(m_1+m_2)\log(m_1^2)\log(m_2^2) \nonumber \\
  q_{0,1}&=& m_1m_2\Bigl[-\frac{1}{6}m_1^3 +\frac{13}{24}m_1^2m_2
      +\frac{13}{24}m_1m_2^2 -\frac{1}{6}m_2^3 \nonumber \\
     &&+\frac{1}{8}m_1m_2(m_1+m_2)(\log(m_1^2)+\log(m_2^2))\Bigr]
  \nonumber \\
  q_{0,2}&=& \frac{1}{4}m_1^2m_2^2(m_1+m_2) \nonumber \\
  q_{1,0}&=& \frac{65}{64}m_1^4 +\frac{1031}{192}m_1^3m_2
     + \frac{857}{96}m_1^2m_2^2 +\frac{1031}{192}m_1m_2^3 +\frac{65}{64}m_2^4
    \nonumber \\
    &&{\kern-25pt}
 -m_1^2 (\frac{3}{16}m_1^2 +\frac{5}{12}m_1m_2 +\frac{7}{16}m_2^2)
               \log(m_1^2)
      -m_2^2 (\frac{7}{16}m_1^2 +\frac{5}{12}m_1m_2 +\frac{3}{16}m_2^2)
               \log(m_2^2) \nonumber \\
  &&-\frac{3}{32}(m_1+m_2)^2\Bigl[m_1^2 \log^2(m_1^2)
    +m_2^2\log^2(m_2^2)\Bigr]
    +\frac{1}{8}m_1^2m_2^2\log(m_1^2)\log(m_2^2) \nonumber \\
 q_{1,1} &=& m_1^4 +\frac{137}{24}m_1^3m_2 +\frac{25}{3}m_1^2m_2^2
                +\frac{137}{24}m_1m_2^3 +m_2^4 \nonumber \\
 && -m_1^2(\frac{3}{8}m_1^2 +\frac{3}{4}m_1m_2+\frac{5}{8}m_2^2)\log(m_1^2)
    -m_2^2(\frac{5}{8}m_1^2 +\frac{3}{4}m_1m_2+\frac{3}{8}m_2^2)\log(m_2^2)
   \nonumber \\
  q_{1,2} &=&-\frac{1}{8}\left(3m_1^4 + 6m_1^3m_2 +10m_1^2m_2^2 +6m_1m_2^3
               +3m_2^4\right) \nonumber \\
  q_{2,0} &=& -(m_1+m_2)\Bigl[ \frac{193}{64}m_1^2 + \frac{1775}{192}m_1m_2
    +\frac{193}{64}m_2^2 \nonumber \\
   &&+ \frac{3}{8} m_1^2\log(m_1^2) + \frac{3}{8} m_2^2\log(m_2^2)
   -\frac{9}{32} m_1^2\log^2(m_1^2)
   -\frac{9}{32} m_2^2\log^2(m_2^2) \Bigr] \nonumber \\
  q_{2,1} &=& (m_1+m_2)\Bigl[ -\frac{63}{8}m_1^2 -\frac{433}{24}m_1m_2
          -\frac{63}{8}m_2^2 +\frac{9}{8}m_1^2\log(m_1^2)
          +\frac{9}{8}m_2^2\log(m_2^2)\Bigr] \nonumber \\
   q_{2,2} &=& \frac{9}{8}(m_1+m_2)(m_1^2+m_2^2) \nonumber \\
  q_{3,0} &=& \frac{251}{64}m_1^2 +\frac{241}{24}m_1m_2 + \frac{251}{64}m_2^2
   \nonumber \\
 &&+\frac{1}{2}m_1^2\log(m_1^2)+\frac{1}{2}m_2^2\log(m_2^2)
   -\frac{3}{16}m_1^2\log^2(m_1^2)
   -\frac{3}{16}m_2^2\log^2(m_2^2) \nonumber \\
  q_{3,1} &=& \frac{153}{8}m_1^2 +39m_1m_2 + \frac{153}{8}m_2^2
   -\frac{3}{4}m_1^2\log(m_1^2)-\frac{3}{4}m_2^2\log(m_2^2)
   \nonumber \\
 q_{3,2} &=&-\frac{3}{4}(m_1^2+m_2^2) \nonumber \\
 q_{4,0} &=& -\frac{37}{32}(m_1+m_2) \ \ \ \
 q_{4,1} = -\frac{79}{4}(m_1+m_2)  \ \ \ \
 q_{5,0} = -\frac{11}{16} \ \ \ \
 q_{5,1} = \frac{15}{2}
\labbel{14} \end{eqnarray}

\def\NP{{\sl Nuc. Phys.}}
\def\PL{{\sl Phys. Lett.}}
\def\PR{{\sl Phys. Rev.}}
\def\PRL{{\sl Phys. Rev. Lett.}}
\def\NC{{\sl Nuovo Cim.}}

\end{document}